\pdfoutput=1

\documentclass[11pt,twoside,a4paper,cmspaper,final,collab]{cms-tdr}
\def\svnVersion{175945}\def\cmsCernNo{2012-304}\def\cmsCernDate{\today}\def\cmsMessage{Submitted to the Journal of High Energy Physics}

\begin{document}\cmsNoteHeader{EXO-11-101}

\hyphenation{had-ron-i-za-tion}
\hyphenation{cal-or-i-me-ter}
\hyphenation{de-vices}

\RCS$Revision: 164854 $
\RCS$HeadURL: svn+ssh://svn.cern.ch/reps/tdr2/papers/EXO-11-101/trunk/EXO-11-101.tex $
\RCS$Id: EXO-11-101.tex 164854 2013-01-15 14:20:37Z tomalin $
\cmsNoteHeader{EXO-11-101} 

\newcommand{\Higgs}{\Hz}
\newcommand{\Pshiggs}{\Az}
\newcommand{\Zprime}{\cPZpr\xspace}
\newcommand{\Kzero}{\ensuremath{\mathrm{K}^0_\mathrm{s}}\xspace}
\newcommand{\lp}{\ensuremath{\ell^+}\xspace}
\newcommand{\lm}{\ensuremath{\ell^-}\xspace}
\newcommand{\X}{\ensuremath{\mathrm{X}}\xspace}
\newcommand{\XX}{\ensuremath{\mathrm{XX}}\xspace}
\newcommand{\Y}{\ensuremath{\mathrm{Y}}\xspace}
\newcommand{\BR}{\ensuremath{\mathcal{B}}\xspace}
\newcommand{\eTrigA}{\texttt{HLT\_\-DoublePhoton33}\xspace}
\newcommand{\eTrigB}{\texttt{HLT\_\-DoublePhoton33\_\-HEVT}\xspace}
\newcommand{\eTrigC}{\texttt{HLT\_\-DoublePhoton38\_\-HEVT}\xspace}
\newcommand{\eTrigD}{\texttt{HLT\_\-DoublePhoton43\_\-HEVT}\xspace}
\newcommand{\muTrigA}{\texttt{HLT\_\-L2DoubleMu23\_\-NoVertex}\xspace}
\newcommand{\muTrigB}{\texttt{HLT\_\-L2DoubleMu30\_\-NoVertex}\xspace}
\newcommand{\muTrigAB}{\texttt{HLT\_\-L2DoubleMu23\!(30)\!\_\-NoVertex}\xspace}

\title{Search in leptonic channels for heavy resonances decaying to long-lived neutral particles}

\date{\today}

\abstract{
A search is performed for heavy resonances decaying to two long-lived massive neutral particles,
each decaying to leptons. The experimental signature is a distinctive topology consisting of a pair
of oppositely charged leptons originating at a separated secondary vertex.  Events were collected by the
CMS detector at the LHC during pp collisions at $\sqrt{s} =7\TeV$, and selected from data samples
corresponding to 4.1 (5.1)\fbinv of integrated luminosity in the electron (muon) channel.
No significant excess is observed above standard model expectations, and an upper limit is set
with 95\%~confidence level on the production cross section times the branching fraction to leptons,
as a function of the long-lived massive neutral particle lifetime.
}

\hypersetup{%
pdfauthor={CMS Collaboration},%
pdftitle={Search in leptonic channels for heavy resonances decaying to long-lived neutral particles},%
pdfsubject={CMS},%
pdfkeywords={CMS, physics}}

\maketitle 

\section{Introduction}

Several models of new physics predict the existence of massive, long-lived particles which could
manifest themselves through their delayed decays to leptons. Such scenarios arise, for example,
in various supersymmetric (SUSY) scenarios such as ``split SUSY''
\cite{Hewett:2004nw} or SUSY with very weak R-parity violation \cite{Barbier:2004ez}, ``hidden valley'' models \cite{Han:2007ae}, and \Zprime models
that contain long-lived neutrinos~\cite{Basso:2008iv}.

This Letter presents the first search using data from the Compact Muon
Solenoid (CMS) for massive, long-lived exotic particles \X that decay
to a pair of oppositely charged leptons. We search for events
containing a pair of oppositely charged
electrons or muons (dileptons) originating from a common secondary
vertex within the volume of the CMS tracker,
that is significantly transversely displaced from the event primary vertex.
These leptons are assumed to originate from a 2-body decay of a long-lived
particle, and so are required to form a narrow resonance in the dilepton mass spectrum.
This topological signature has the potential to provide clear evidence for
physics beyond the standard model (SM). It is also very powerful in suppressing backgrounds from
standard model processes.

This signature is sensitive to a wide class of models.  However, for the
purpose of establishing a signal benchmark, a specific model of a long-lived, spinless,
exotic particle \X which has a non-zero branching fraction to dileptons is used. In this particular
model, the \X is pair-produced in the decay of a Higgs boson, i.e.  $\Higgs\rightarrow
2\X$, $\X\rightarrow\lp\lm$ \cite{Strassler:2006ri}, where the Higgs boson is produced through gluon-gluon
fusion. This model predicts up to two displaced
dilepton vertices in the tracking volume per event.

The \DZERO Collaboration has performed searches for leptons from delayed decays in its
tracker volume \cite{Abazov:2006as,Abazov:2008zm}, but these searches
are sensitive to a much smaller kinematic phase space region than CMS.
The ATLAS Collaboration has performed searches that are sensitive to decay lengths up to
about 20\unit{m} by exploiting the ATLAS muon spectrometer \cite{ATLAS:2012av,Aad:2011zb}, using different
decay channels from those considered in this Letter.

\section{The CMS detector}

The central feature of the CMS apparatus~\cite{JINST} is a
superconducting solenoid of 6\unit{m} internal diameter providing an axial field
of 3.8\unit{T}. Within the field volume are the silicon pixel and strip
tracker, the lead-tungstate crystal electromagnetic calorimeter (ECAL),
and the brass/scintillator hadron calorimeter. Muons are identified in
gas-ionisation detectors embedded in the steel magnetic-flux return yoke of the
solenoid.

The silicon tracker is composed of pixel detectors (three barrel layers and two forward disks on either
end of the detector) surrounded by strip detectors (ten barrel layers plus three inner disks and nine
forward disks at each end of the detector). The tracker covers the pseudorapidity range $|\eta| < 2.5$,
where $\eta = -\ln[\tan(\theta/2)]$ and $\theta$ is the polar angle with respect
to the anticlockwise-beam direction. All tracker layers provide two-dimensional hit position measurements, but only the pixel
tracker and a subset of the strip tracker layers provide three-dimensional hit position
measurements. Owing to the strong magnetic field and the high granularity of the silicon tracker,
promptly produced charged particles with transverse momentum $\pt = 100\GeVc$ are reconstructed with a
resolution in \pt of ${\approx}1.5\%$ and in transverse impact parameter $d_0$ of ${\approx}15\mum$.  The
track reconstruction algorithms are able to reconstruct displaced tracks with transverse impact
parameters up to ${\approx}25\unit{cm}$ from particles decaying up to ${\approx}50\unit{cm}$ from the beam line.  The
performance of the track reconstruction algorithms has been studied with data~\cite{Khachatryan:2010pw}.The silicon
tracker is also used to reconstruct the primary vertex position with a
precision of $\sigma_d\approx 20\mum$ in each dimension.

The ECAL consists of nearly 76\,000 lead
tungstate crystals, which provide coverage in pseudorapidity $|\eta| < 3$.
Muons are measured in the pseudorapidity range $|\eta|< 2.4$ with
detection planes based on one of three technologies: drift tubes in the barrel
region, cathode strip chambers in the endcaps, and resistive plate chambers
in the barrel and endcaps.

The first level of the CMS trigger system, composed of custom hardware processors, selects events of
interest using information from the calorimeters and the muon detectors. A
high-level trigger processor farm then employs the full event information to
further decrease the event rate.

\section{Data and Monte Carlo simulation samples}
\label{sec:samples}

For this analysis, pp collision data at a centre-of-mass energy of 7\TeV corresponding to an integrated
luminosity of $4.1 \pm 0.1$ ($5.1 \pm 0.1$)\fbinv are used in electron (muon) channels. (The lower electron
luminosity is due to the fact that not all data was taken with triggers suitable for this analysis.)

For the electron channel, these data are collected with a trigger that requires
two clustered energy deposits in the ECAL, each with
transverse energy $E_T > 38$\GeV. For the muon channel, the trigger requires two muons,
each reconstructed without using any primary vertex constraint and having $\pt > 30$\GeVc.
The tracker information is not used in either trigger.

Signal samples are generated using \PYTHIA V6.424 \cite{PYTHIA} to
simulate \Higgs~production through gluon fusion ($\Pg\Pg \rightarrow\Higgs$).  Subsequently the \Higgs~is forced
to decay to two long-lived spin~0 exotic particles ($\Higgs\rightarrow\X\X$), which then decay to dileptons
($\X\rightarrow\lp\lm$), where $\ell $ represents either a muon or an electron. Several samples with different
combinations of \Higgs~masses ($M_{\Higgs}$ = 125, 200, 400, 1000\GeVcc ) and \X~boson masses ($M_{\X}$ = 20,
50, 150, 350\GeVcc) are generated. The lifetimes of \X~ bosons used in these samples are chosen to give a
mean transverse decay length of approximately 20\cm in the laboratory frame.  Several simulated background
samples generated with \PYTHIA are used, corresponding to $\ttbar$, $\Z/\gamma\rightarrow\lp\lm$ (including
jets), W/\Z boson pair production with leptonic decays, and QCD multijet events. The contribution from single
W + jet production is negligible. In all the samples, the response of the detector is simulated
in detail using \GEANTfour~\cite{GEANT4}. The samples are then processed through the trigger emulation and
event reconstruction chain of the CMS experiment.

\section{Event reconstruction and selection}
\label{sec:RecoAndSelection}

Events are required to contain a primary vertex, which has at
least four associated tracks and whose position is displaced from the nominal
interaction point by no more than 2\unit{cm} in the direction transverse to the beam
and no more than 24\unit{cm} in the direction along the beam. Furthermore, to reject
events produced by the interaction of beam-related protons with the
LHC collimators, in any event with at least 10 tracks
(counting all tracks irrespective of whether they are associated with a primary vertex) 
the fraction of the tracks that are classified as ``high purity'', as
defined in Ref.~\cite{Khachatryan:2010pw}, must exceed 25\%. This requirement is not imposed on
events with less than 10 tracks.

The selection of lepton candidates from displaced secondary vertices begins by searching for high-purity
tracks with transverse momenta $\pt > 41\GeVc$ (33\GeVc) for the electron
(muon) channel.   These  criteria are slightly higher
than the corresponding trigger thresholds, to minimise dependence on the trigger
inefficiency in the $\pt$ turn-on region.
The tracks are required to have pseudorapidity $|\eta|<2$, as the efficiency for
finding tracks from displaced secondary vertices falls off at large $|\eta|$. To reject
promptly produced particles, the tracks must have a transverse impact parameter
significance with respect to the beam line of $|d_0/\sigma_d| > 3$ (2) in the
electron (muon) channel.
Because bremsstrahlung in the material of the tracker significantly
affects the reconstruction of electrons, the electron channel selection
criteria are more restrictive.
Tracks are considered to be
identified as leptons if $\Delta R = \sqrt{(\Delta\eta)^2+(\Delta\phi)^2}$ is less than 0.1.
Here, $\Delta\eta$ and $\Delta\phi$ are the differences between the track and a lepton trigger object
in pseudorapidity and $\phi$, the azimuthal angle about the anticlockwise-beam direction.
Standard CMS offline lepton identification
algorithms are not applied, since they are inefficient for leptons
from highly displaced vertices. However these algorithms are not needed to suppress the very low backgrounds
present in this analysis. For the electron channel, the energy of the electron
is estimated from a deposit in the ECAL that is near the reconstructed track
trajectory.

The \X~boson candidates are formed from pairs of oppositely-charged
lepton candidates.  The two corresponding tracks are fitted to a common vertex,
which is required to have a chisquare per degree of freedom
$\chi^2/\text{dof} < 5$, where $\text{dof} = 1$.  For events in the electron (muon) channel,
the vertex must lie at a distance of more than 8 (5) standard
deviations from the primary vertex in the transverse plane.
If either track has more than one hit closer to the centre of CMS than
their common vertex, the event is rejected.

Both lepton candidates are required to be isolated, to reject
background from jets. A hollow isolation cone of radius $0.03 < \Delta R < 0.3$
is constructed around
each candidate. Within this isolation cone, the $\sum\pt$ of all
tracks with $\pt > 1$\GeVc, excluding the other lepton candidate, must be less than 4\GeVc.
This requirement has very little effect on the signal efficiency,
which is relatively insensitive to the number of primary vertices in
each event.
According to simulation, the mean $\sum\pt$ in the
isolation cone increases from 0.6 to 1.2\GeVc
as the number of additional primary vertices per event increases
from 0 to 20.

Cosmic ray muons may be reconstructed as back-to-back tracks. To reject them, a requirement of $\cos(\alpha) >
-0.95$ is applied, where $\alpha$ is the opening angle between the two tracks.  Background from misidentified
leptons is reduced by requiring that the two lepton candidates are not both matched to the same trigger
object. Additional background rejection is achieved by requiring that, projected into the plane perpendicular
to the beam line, the reconstructed momentum vector of the \X boson candidate is collinear with the vector
from the primary vertex to the secondary vertex. The collinearity angle is required to be less than 0.8 (0.2)
radians in the electron (muon) channel.  Owing to the difficulty of modelling the trigger efficiency for closely
spaced muon pairs, the two tracks in muon channel candidates must be separated by $\Delta R > 0.2$.  To
eliminate background from \JPsi and $\Upsilon$ decays and from $\gamma$ conversions, X boson candidates are
required to have dilepton invariant masses larger than 15\GeVcc.  If more than one \X~boson candidate is
identified in a given event, all the selected candidates are retained.

\subsection{Selection efficiency}
\label{sec:effi}

The selection efficiency for the electron (muon) channel is defined as the fraction of the $\X\rightarrow e^+e^-$ ($\mu^+\mu^-$) 
decays that pass the full set of selection criteria, as evaluated using the simulated signal events. It is evaluated separately for
two different classes of events:
first for $\Higgs\rightarrow\X\X$ events in which only one long-lived exotic particle decays to the chosen
lepton species, defining efficiency $\epsilon_{1}$, and second for events in which both long-lived
exotic particles decay to chosen lepton pairs, defining efficiency $\epsilon_{2}$.  The efficiencies $\epsilon_{1}$ and
$\epsilon_{2}$ are usually almost identical, indicating that the efficiency to select an \X~boson candidate is
not strongly affected by whether or not the second \X~boson in the event decays to the same lepton channel.
The only exception is for the muon channel in the case of small $M_\X / M_{\Higgs}$,
where the dimuon trigger is inefficient for the two nearly collinear muons from the decay of the same
\X~boson, but the trigger requirement can still be satisfied by muons from separate \X~bosons.
The efficiencies are estimated for a range of \X~boson lifetimes, corresponding to mean transverse
decay lengths of $\approx 0.7$ -- 600\unit{cm}, by reweighting the simulated signal events.
The maximum efficiency (for $M_{\Higgs} = 1000\GeVcc$, $M_X = 150\GeVcc$, $c\tau = 1\unit{cm}$) is approximately
34\% (52\%) in the electron (muon) channel, but becomes significantly smaller at lower \Higgs~masses or longer
lifetimes.

\section{Background estimation and modelling}
\label{sec:bkgndest}

An interpretation of the observed dilepton mass spectrum requires an estimate of the background normalisation
and a parametrisation of the background shape as a function of $M_X$.

\subsection{Background normalisation}
\label{sec:bkgndnorm}

\begin{figure*}[hbtp]
  \begin{center}
\includegraphics[width=0.45\textwidth]{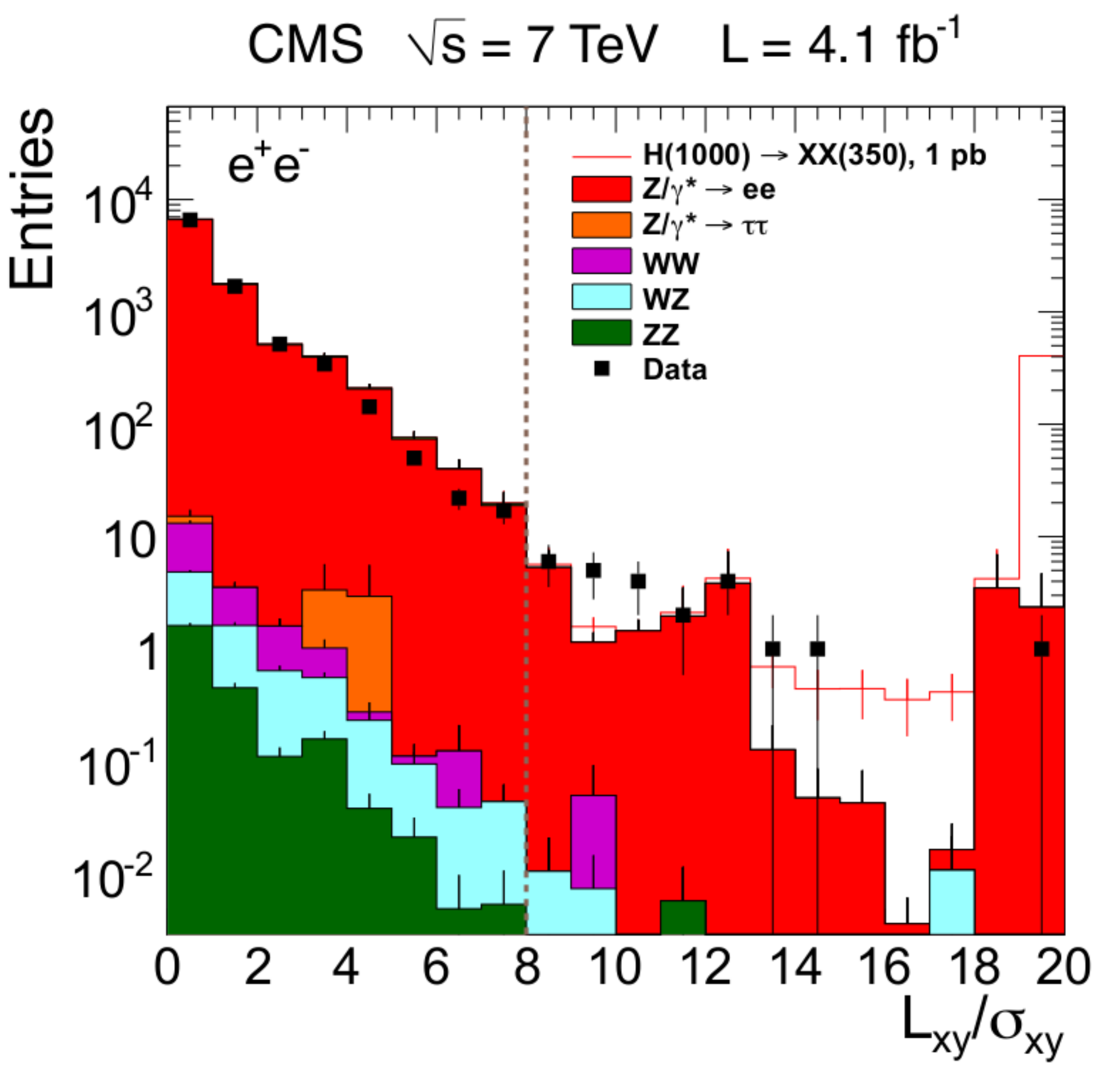}
    \hspace{1cm}
    \includegraphics[width=0.45\textwidth]{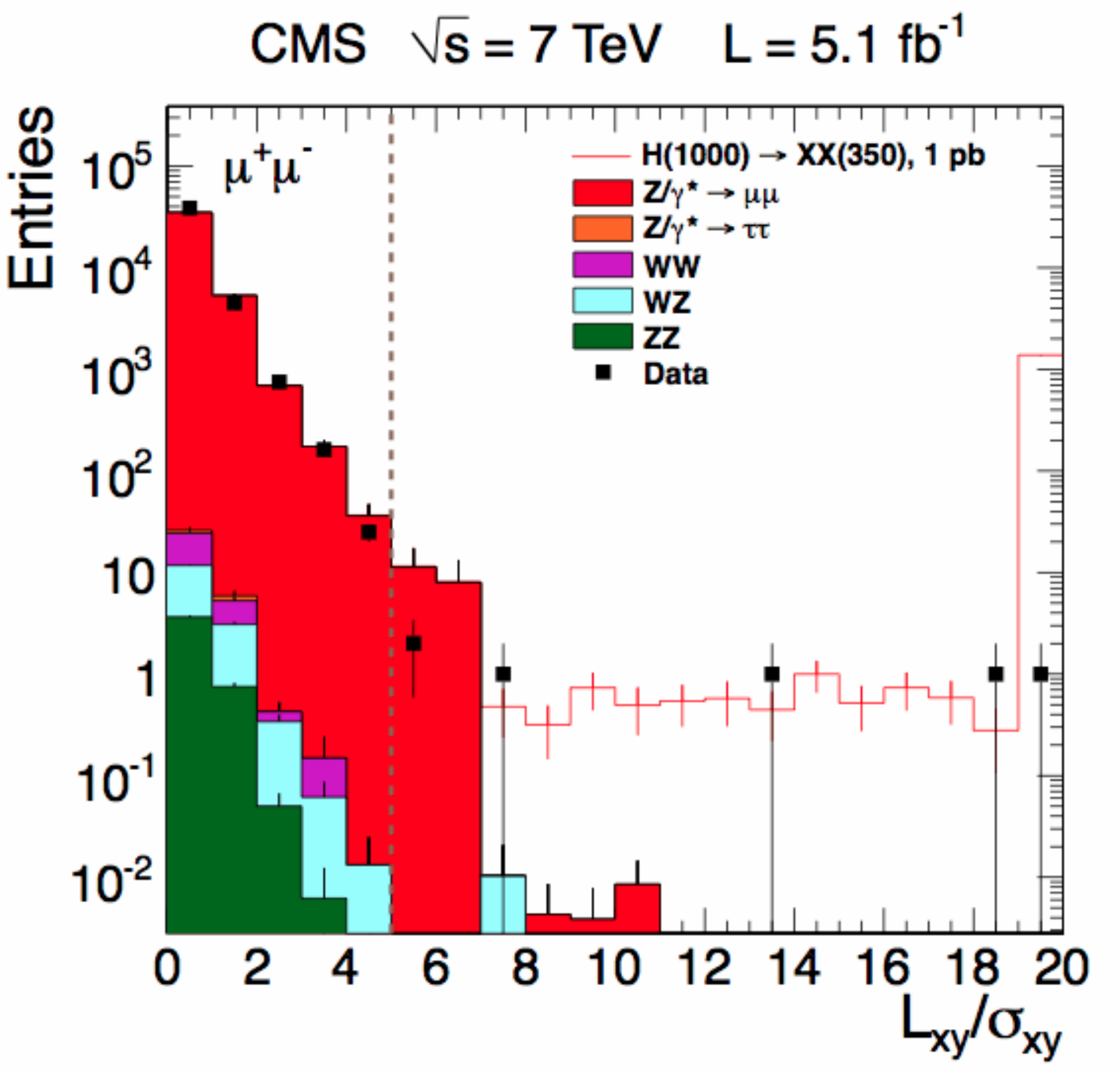}
    \caption{The transverse decay length significance of the candidates
    for the dielectron (left) and dimuon (right) channels with loosened cuts in data and simulation. The
    vertical dashed line indicates the selection requirement used for signal events. There are no simulated
    QCD or $\ttbar$ events passing these selection requirements, so they are omitted.}
    \label{fig:BackgroundFits}
  \end{center}
\end{figure*}

The number of background events passing all the selection criteria for \X boson candidates, is estimated from
simulated samples using the distribution of the transverse decay length significance $L_{xy}/\sigma_{xy}$. We parameterize this distribution
with the sum of two falling exponentials. By integrating the
fitted curve over the signal region, defined by $L_{xy}/\sigma_{xy} >$ 8 (5) for the electron (muon)
channel, an estimate of the mean total background in the mass spectrum is obtained. The estimate gives
$1.4^{+1.8}_{-1.2}$ ($0.02^{+0.09}_{-0.02}$) candidates in the electron (muon) channel. This estimate of
the mean total background is used to derive the results in Section~\ref{sec:results}.
To verify that the simulation correctly describes the $L_{xy}/\sigma_{xy}$ spectrum, data and simulation distributions
are compared in Fig.~\ref{fig:BackgroundFits}, after removing the lifetime-related selection requirements to increase the number 
of events and ensure that the plots are dominated by background.
Specifically, the thresholds on the decay length significance $L_{xy}/\sigma_{xy}$ and individual lepton
$d_0/\sigma_{d}$ were removed entirely.

Figure~\ref{fig:BackgroundFits} also shows that the main background to this search consists
 of prompt dileptons that have been reconstructed with large decay length significance.

\subsection{Background shape}
\label{sec:bkgndshape}

\begin{figure*}[hbtp]
  \begin{center}
    \includegraphics[width=0.45\textwidth]{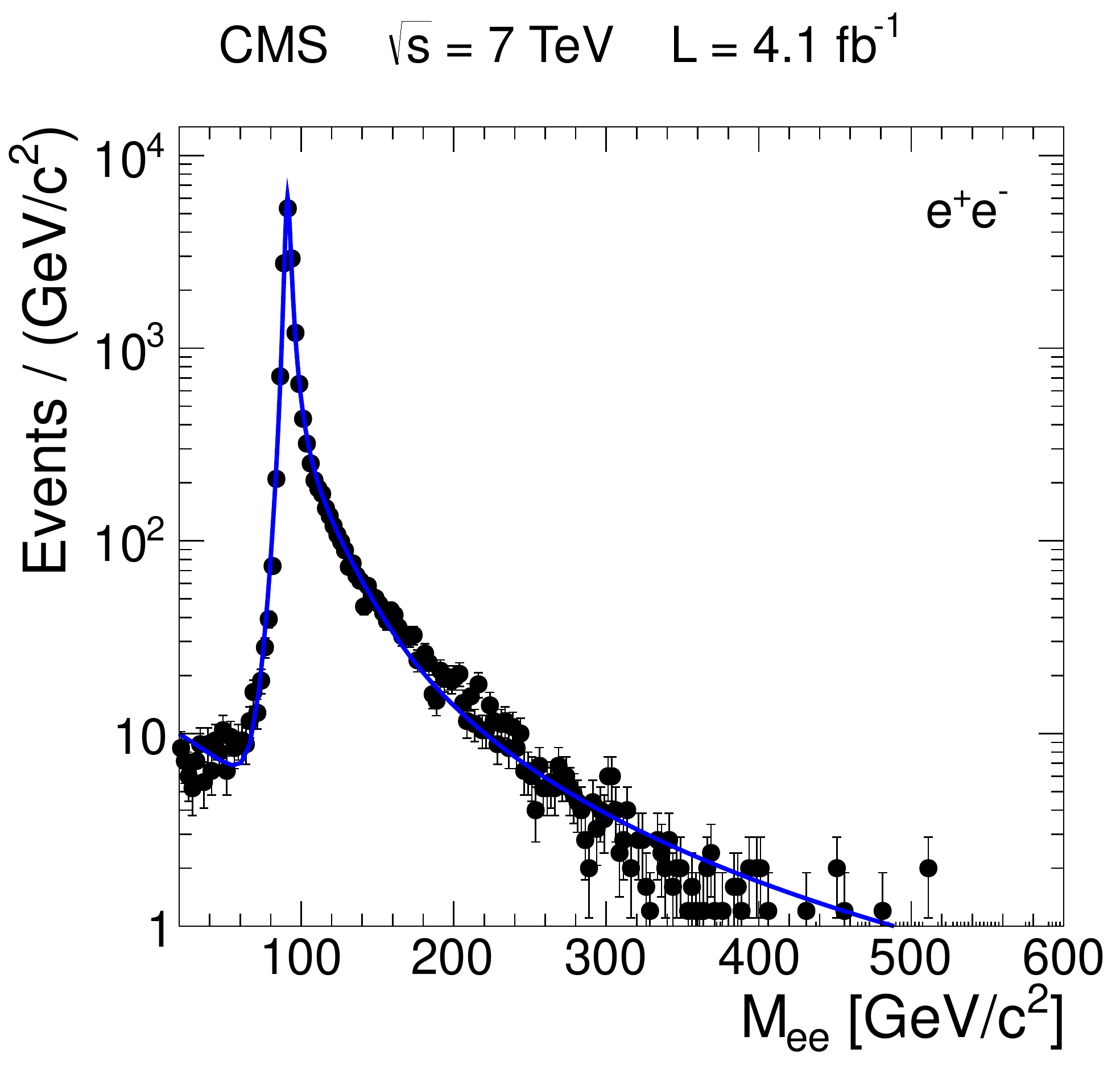}
      \hspace{1cm}
      \includegraphics[width=0.45\textwidth]{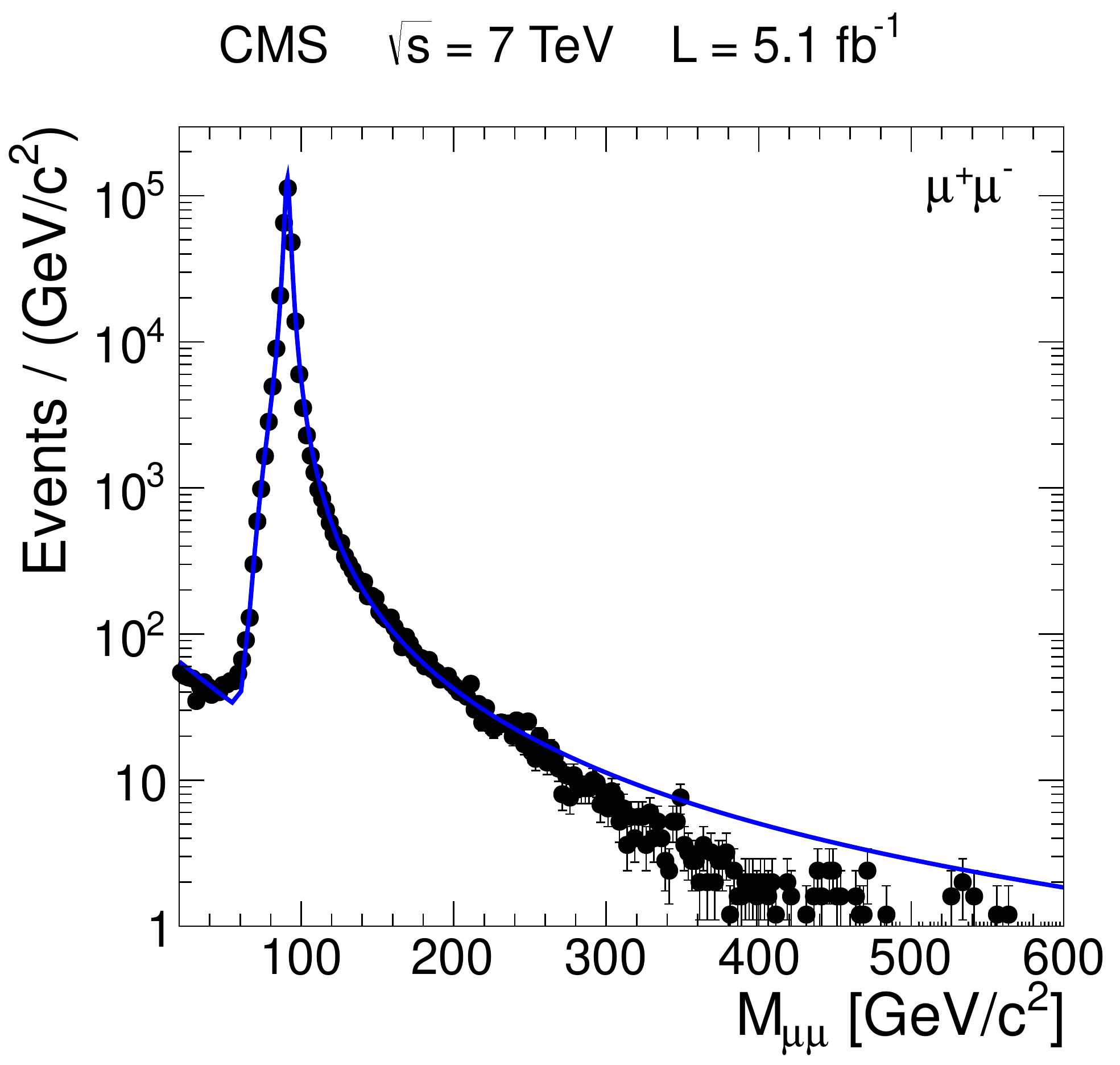}
    \caption{Distribution of the dilepton mass and the fitted shape in a data sample with
    lifetime-related selection requirements removed, shown for the electron (left) and muon (right)
    channels. The shape used is that of a Breit--Wigner distribution times a turn-on function, added to an
    exponential term.}
    \label{fig:BackgroundPDF}
  \end{center}
\end{figure*}

An estimate of the background shape can be obtained from the $M_X$ distribution of a background
sample. However, after applying all selection requirements there are too few events to measure its
shape accurately. Since the dilepton mass distribution and lifetime-related variables are only weakly
correlated in simulated background candidates, the shape of the mass distribution is
instead obtained by fitting a parameterized function to
data samples with the lifetime-related selection requirements
removed. Namely, no selections are made on the individual lepton
$d_0/\sigma_{d}$,
the transverse decay length significance $L_{xy}/\sigma_{xy}$, or the collinearity angle $\Delta\varphi$.
Figure~\ref{fig:BackgroundPDF} shows the results of these fits to the electron and muon data samples.
The observed background is approximately described by the sum
of two functions: the first being a Breit--Wigner function, to represent the \Z~resonance, multiplied by a Gaussian error
function, to approximate the effect on the selection efficiency of the lepton \pt thresholds;
and the second being a falling exponential function, to represent the non-\Z background. The fits give the fraction of the
background from the Z as $0.985 \pm 0.002$ ($0.994 \pm 0.001$) for the
electron (muon) channels.

\section{Systematic uncertainties}
\label{sec:Systematics}

The primary systematic uncertainty comes from the efficiency in detecting and reconstructing signal
events. This uncertainty derives from uncertainties in the efficiency of reconstructing tracks
from displaced vertices,
the trigger efficiency, the modelling of pileup in the simulation, the parton distribution function sets,
the renormalisation and factorisation scales used in generating simulated events, and the effect
of higher order QCD corrections. In addition, systematic
uncertainties in the integrated luminosity, and the background estimate are considered.

Table~\ref{tab:syst} summarises the sources of systematic uncertainty affecting the signal efficiency.
The relative uncertainty in the luminosity is taken to be 2.2\% \cite{cms:lumi11WinterUpdate}.

\begin{table*}[hbtp]
\begin{center}
\topcaption{Systematic uncertainties affecting the signal efficiency over
  the range of $M_{\Higgs}$ and $M_{\X}$ values considered. In all cases, the uncertainty specified is
  a relative uncertainty. Note that the NLO uncertainty is only evaluated for the $M_{\Higgs} = 125$\GeVcc
  case. The relative uncertainty in the luminosity is taken to be 2.2\%.}
\label{tab:syst}
\begin{tabular}{r|l}
Source & Uncertainty \\
\hline
Pileup modelling & $2\%$ \\
Parton distribution functions & ${<}1\%$ \\
Renormalisation and factorisation scales & ${<}0.5\%$ \\
Tracking efficiency & $20\%$ \\
Trigger efficiency & $2.6\%$ ($\Pe$), $11\%$ ($\mu$) \\
NLO effects ($M_{\Higgs} = 125$\GeVcc only) & $4$--$12\%$ \\
\end{tabular}
\end{center}
\end{table*}

Varying the modelling of the pileup within its estimated uncertainties yields a relative change in
the signal selection efficiency of less than 2\%, irrespective of the mass point chosen.
The relative uncertainty due to parton distribution functions (PDF) is studied
using the PDF4LHC procedure \cite{Bourilkov:2006cj} and is less than
1\% for all mass points.
The dependence of the acceptance on the choice of the renormalisation and factorisation scales, which are
chosen to be equal and are varied by factors of 0.5 and 2, is found to be well below 0.5\%. These
uncertainties are applied in the cross-section limit calculation.

\subsection{Track-finding efficiency}

Understanding the efficiency to find a track as a function of its impact parameter is a crucial aspect of
the analysis. Two methods are used to assess if the efficiency to reconstruct displaced tracks is correctly
modelled by the simulation.

\begin{figure*}[hbtp]
  \centering
  \includegraphics[width =0.45\textwidth]{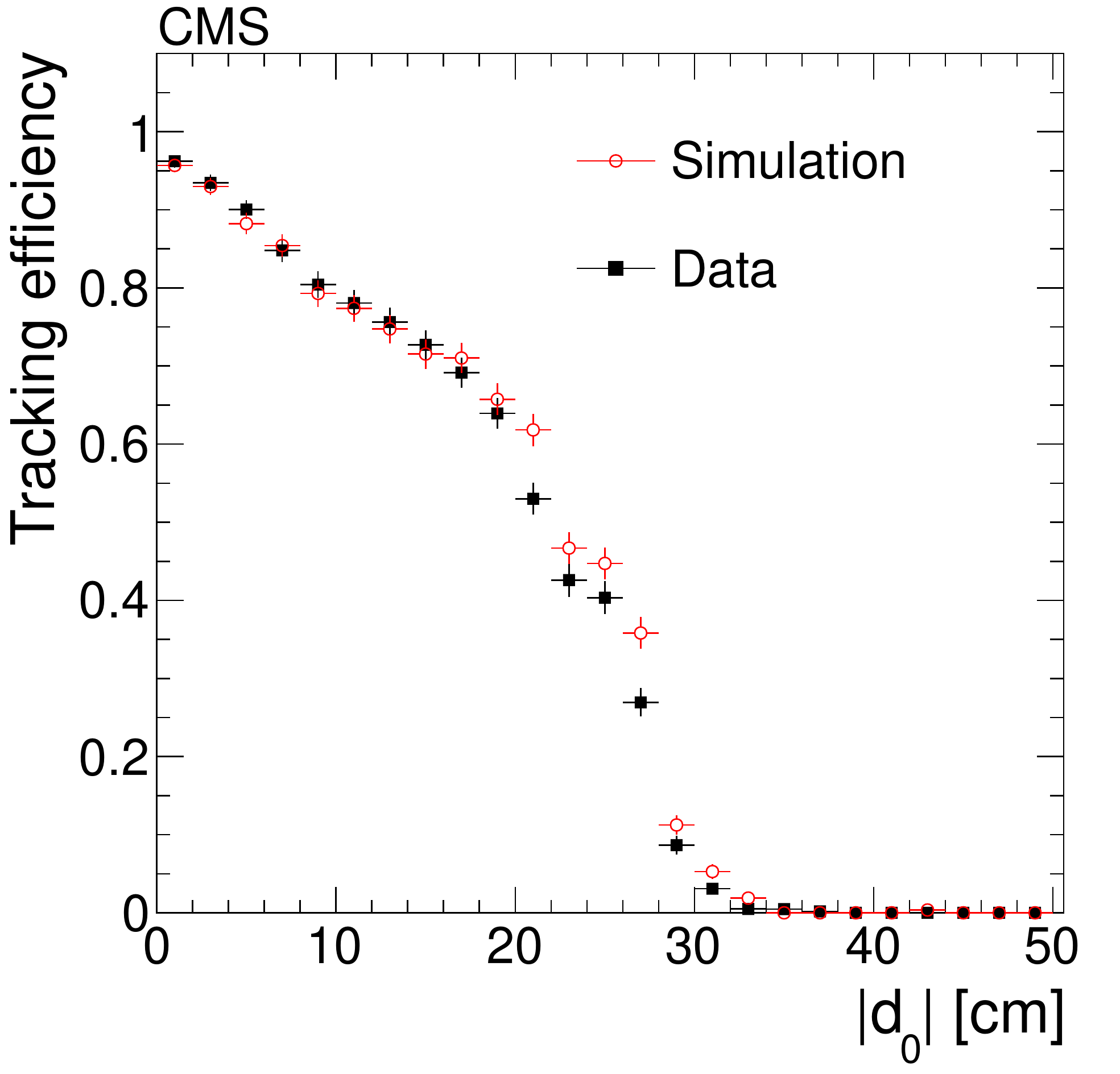}
  \hspace{1cm}\includegraphics[width =0.45\textwidth] {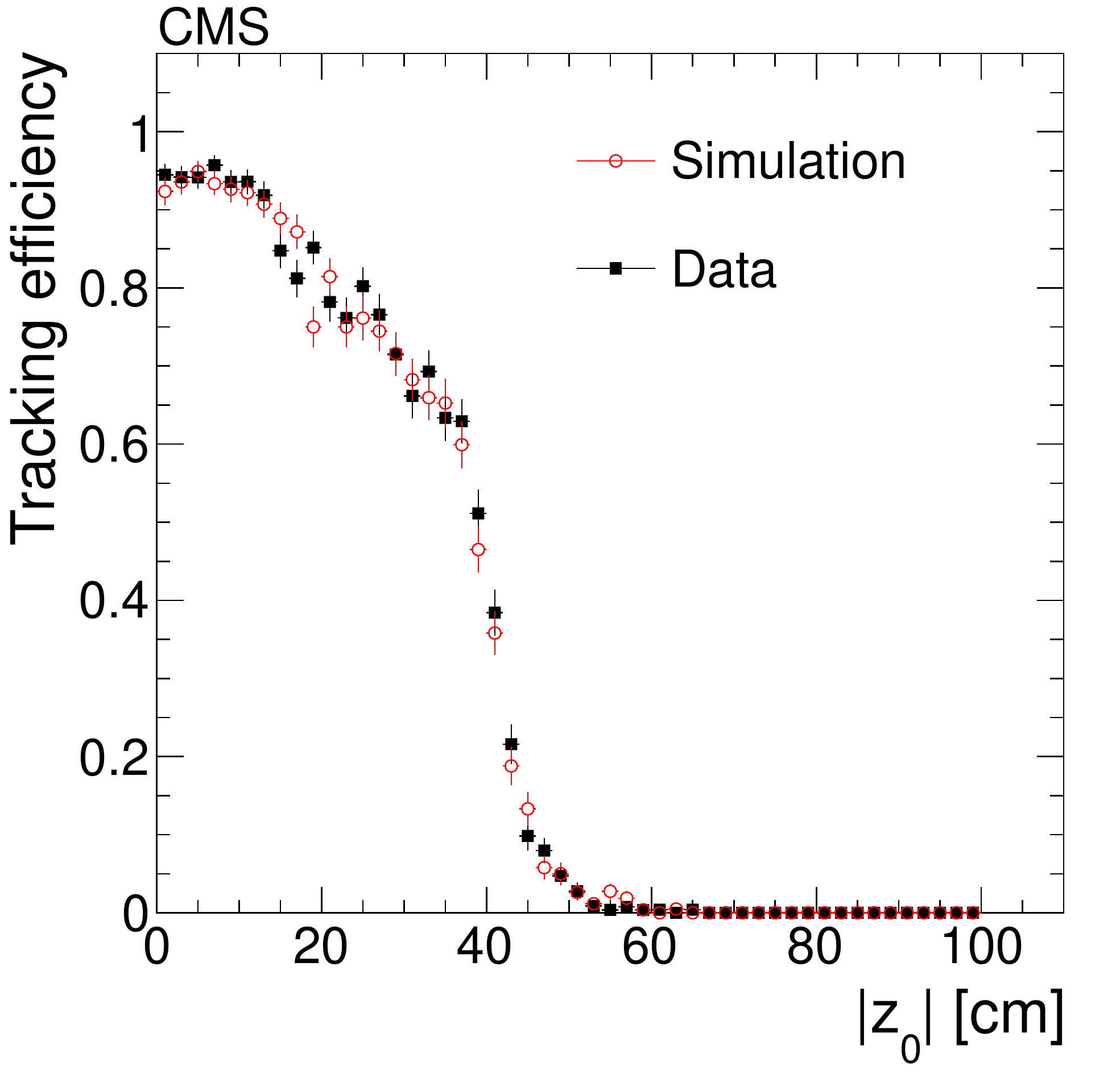}
  \caption{Efficiency of the tracker to find a track, given a
    cosmic ray muon reconstructed in the muon chambers, as a function of the
    transverse (left) and longitudinal (right) impact parameters (with
    respect to the nominal interaction point of CMS). The efficiency
    is plotted in bins of 2\unit{cm} width.
    For the left plot, the longitudinal impact parameter $|z_0|$ is required to be less than
    10\unit{cm}, and for the right plot, the transverse impact parameter $|d_0|$ must be less than
    4\unit{cm}.}
  \label{fig:EffVsImpact}
\end{figure*}

The first method consists of a direct measurement of the efficiency to reconstruct isolated tracks, using
cosmic ray muons. Events are selected from dedicated runs with no beam activity and the cosmic ray muons
are reconstructed by combining the hits in the muon chambers from opposite halves of CMS.  The efficiency
to reconstruct a tracker track associated with a cosmic ray muon, as a function of the transverse and
longitudinal impact parameters, is shown in Fig.~\ref{fig:EffVsImpact}. We focus principally on the
region $|d_0| < 20\unit{cm}$, since in simulated signal, the reconstructed tracks from displaced vertices lie predominantly
in this region. Data and simulation agree within 10\% in this region, so the corresponding relative
systematic uncertainty in the efficiency to reconstruct dilepton candidates is thus taken to be 20\%, as there
are two tracks per candidate.
This method does not explicitly measure tracking efficiency for dielectron candidates. However, simulation
studies indicate that the electron efficiency is only about 10\% smaller than the muon efficiency,
where the difference can be attributed to bremsstrahlung in the electron case. The material budget in
the tracker is modelled in simulation with an accuracy better than 10\% \cite{pas-trk-10-003}. It is
assumed that the difference in tracking efficiency between electrons and muons is modelled with similar
precision. The difference can therefore be neglected in comparison with the much larger systematic
uncertainty in the efficiency to reconstruct dilepton candidates.

To determine if the simulation properly describes the tracking efficiency in the presence of a high density of hits,
 a second method is used.
Single charged particles are simulated at various production points throughout the tracker volume. These particles are then embedded in
both data and simulated data events, and the difference in efficiency to reconstruct the particles
is compared. The relative difference between data and simulation is
less than 1\%.

\subsection{Trigger efficiency measurement}
\label{sec:trigEffi}

The trigger efficiency is measured using the ``tag-and-probe''
method~\cite{Khachatryan:2010xn}.
The decays of \Z~bosons to dileptons are reconstructed in data collected with
single-lepton triggers. They are used to measure the efficiency
for a lepton to pass one leg of the dilepton triggers used in this analysis.
The dilepton trigger efficiency is then obtained as the square
of this single-lepton efficiency, which assumes that there is no correlation
in efficiency between the two leptons.
This is generally a good assumption except in the muon channel, where
dimuons separated by $\Delta R < 0.2$ must be excluded because the trigger is inefficient
for closely spaced dimuons.
The systematic uncertainty associated with the trigger efficiency is evaluated by taking the
difference between the estimates of the efficiency from data and simulation,
yielding a total relative uncertainty of $2.6\%$ for the electron channel and $11\%$ for
the muon channel.

\subsection{Effect of higher-order QCD corrections}
\label{sec:NLOcorr}
For $M_{\Higgs} = 125\GeVcc$, the leptons from the \X~boson decay have a combined efficiency of only a few
percent for passing the lepton \pt requirements. For this reason the signal efficiency at this mass is sensitive
to the modelling of the Higgs \pt spectrum, which may in turn be influenced by higher order QCD corrections.
To study this effect, we reweight the LO \Higgs \pt spectrum from our signal sample to match
the corresponding Higgs $\pt$ spectrum evaluated at NLO. For $M_{\Higgs} = 125\GeVcc$ and
$M_{X} = 20~(50)\GeVcc$ the signal efficiency changes by 4\% (12\%). This
change is taken as an
additional systematic uncertainty in the efficiency for the $M_{\Higgs} = 125\GeVcc$ case. For larger
\Higgs masses, the corresponding systematic uncertainty is below
0.5\%, and hence neglected.

\subsection{Background uncertainty}
\label{sec:bkgsyst}

The systematic uncertainty for the background normalisation is taken from the uncertainty on the
background fits described in Section~\ref{sec:bkgndnorm}, where these fits yield
$1.4^{+1.8}_{-1.2} (0.02^{+0.09}_{-0.02})$ candidates in the electron (muon) channel.
In addition, a cross-check is made by repeating
the fit described in Section~\ref{sec:bkgndnorm}, but restricting the fit to the background region
$L_{xy}/\sigma_{xy} <$ 8 (5) in the electron (muon) channel and then extrapolating it into the signal region. This cross-check is performed for both
simulated events and data. Additionally, the background normalisation is estimated by simply
counting the number of candidates in the simulated background passing all selection criteria.
All methods give consistent results.
As explained in Section~\ref{sec:results}, the uncertainty in the
background normalisation actually affects the observed limits only for \X~boson masses
close to the \Z~resonance.

To determine the sensitivity of the limits to variations in the assumed background shape, described in
Section~\ref{sec:bkgndshape}, other functional forms are also tried. The resulting effect on the
limits, which is taken as a systematic uncertainty, is negligibly small.
The fit for the background shape is repeated using simulated events and consistent results are obtained.  As an
additional check that the shape is not strongly influenced by the lifetime-related requirements, the mass
distribution obtained by relaxing the lifetime-related selection requirements is also fitted, in both data and
simulated events. In all cases, the resulting change in the limits is negligible.

\section{Results}
\label{sec:results}

After all selection requirements are applied, no candidates survive in the muon channel,
consistent with the expected mean number of  $0.02^{+0.09}_{-0.02}$.  In the electron channel,
a total of 4 candidates remain, which is also consistent with the expected mean number of
$1.4^{+1.8}_{-1.2}$. However, 2 of these candidates fall in the region above the \Z mass,
where the estimated background is lower. The observed candidates all have transverse decay
lengths less than 0.3\unit{cm}.

We set 95\% confidence level
(CL) upper limits on the signal process using the CL$_\mathrm{s}$ method \cite{Read:2002hq, Junk:1999kv}, which
makes use of an unbinned likelihood fit to the dilepton mass spectrum.

This fit to the mass spectrum uses the following functions:

\begin{itemize}
\item A Gaussian signal function to represent the signal's mass distribution. For each
\Higgs~mass, the mass resolution used in the Gaussian is obtained from the simulated signal
samples, as a function of the \X~boson mass and lifetime, interpolating between the generated \X~boson masses
with a smooth curve when necessary.
\item The sum of two background functions, one distribution representing the background from the \Z~peak and another more
slowly varying distribution representing the non-\Z background. These
functions are obtained as described in
Section~\ref{sec:bkgndest}.
\end{itemize}

The limit calculation takes into account the systematic uncertainties described in
Section~\ref{sec:Systematics} by introducing a nuisance parameter for each uncertainty, marginalized by a
log-normal prior distribution. In particular, the normalisation of the \Z (non-\Z) background can be
constrained by the a priori estimate of the total background normalisation presented in
Section~\ref{sec:bkgndnorm}, multiplied by the estimated \Z (non-\Z) background fraction from
Section~\ref{sec:bkgndshape}.  However, the fits to the mass spectrum performed as part of the limit
calculation strongly constrain the normalisation of the non-\Z background, even though the normalisation of a
signal is unknown. As a result, it is not necessary to use the a priori estimate of the non-\Z background
normalisation when calculating the observed limits. In the case of the \Z
background, the a priori estimate of the normalisation  is used, but it affects only the limits for \X~bosons whose mass is close to
that of the \Z.  To calculate expected limits, one must have a prediction of the background normalisation,
which is taken from Section~\ref{sec:bkgndest}.

\begin{figure*}[hbtp]
\centering
\includegraphics[width=0.49\textwidth]{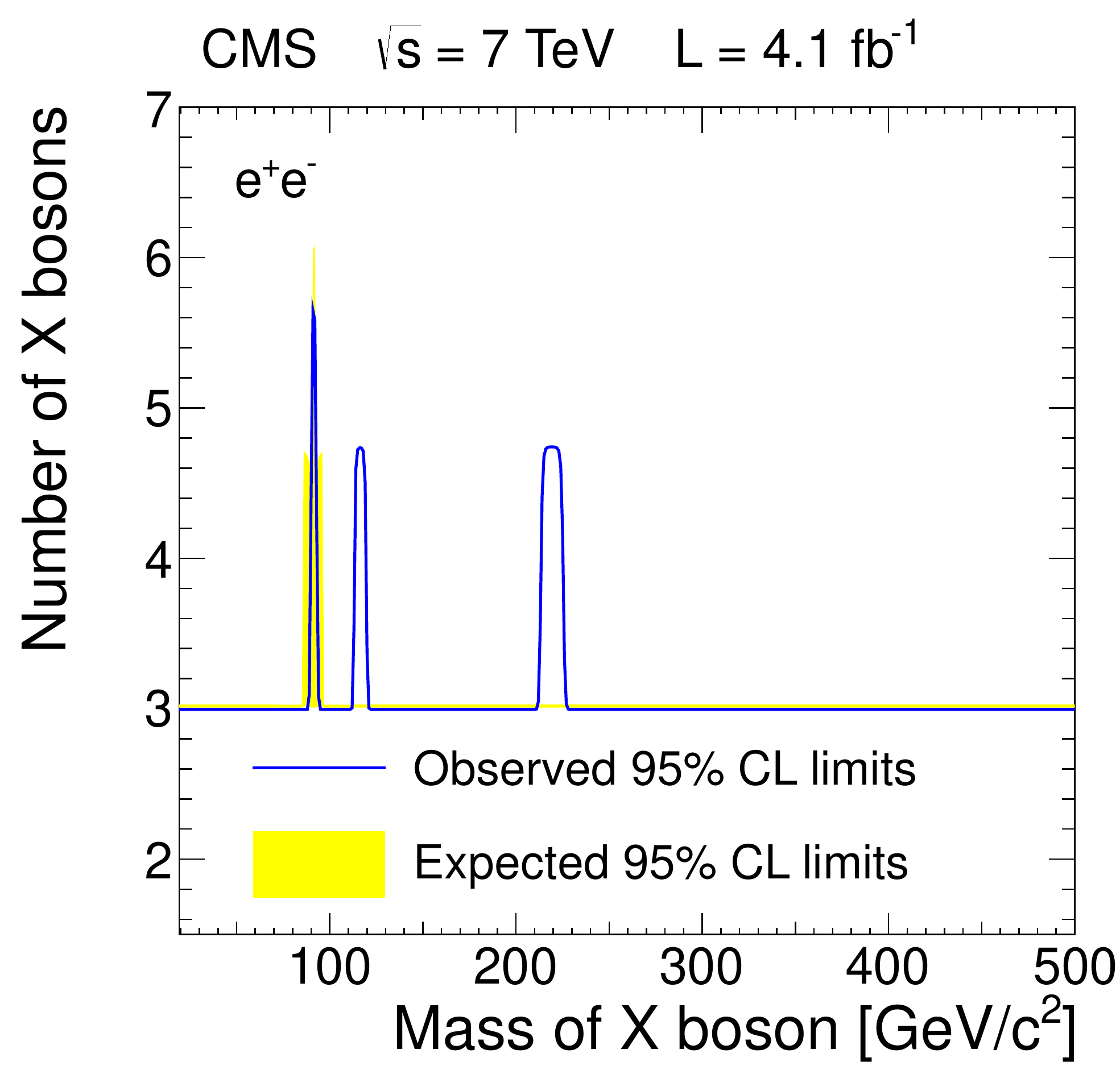}
\includegraphics[width=0.49\textwidth]{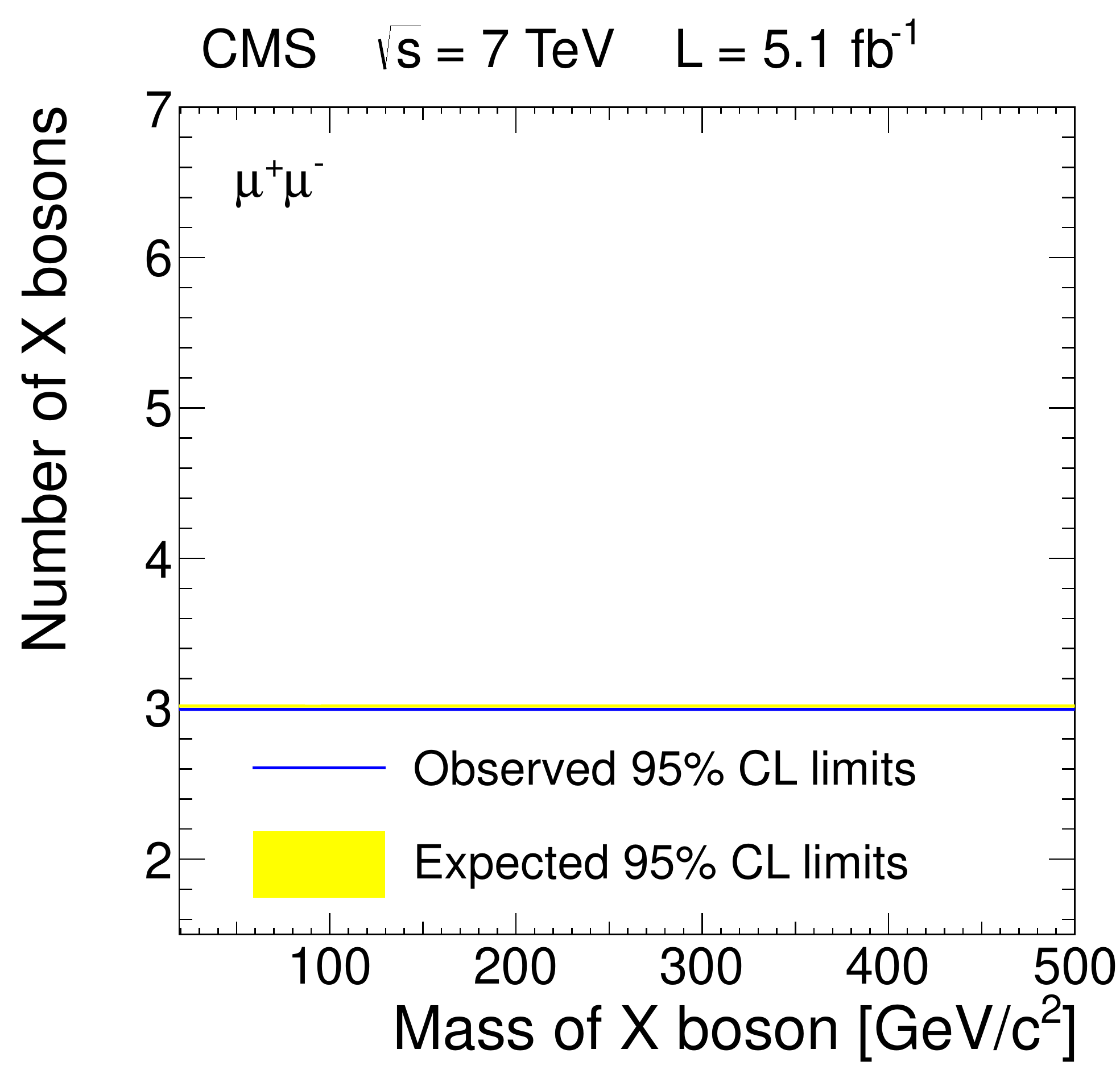}
\caption{The 95\% CL upper limits on the mean number of \X~bosons that could pass the
selection requirements in the electron (muon) channels are shown in the left (right)
plot. A yellow shaded band shows the
95\% quantile for the expected limits, but is almost entirely hidden by the observed limit curves.}
\label{fig:FinLimitsNum}
\end{figure*}

As a first step, upper limits are placed on the mean number $N_X$ of \X~bosons that could pass the
selection requirements, as a function of the \X~boson mass.
The resulting upper limits on $N_X$ at 95\% CL for the electron and muon channels are presented in
Fig.~\ref{fig:FinLimitsNum}. These limits are independent of the particular model assumed for \X~boson
production, except for the mass resolution assumed for the signal, which affects the width of the
resulting peaks in the observed limit. The mass resolution used for these limits is derived from the
Monte Carlo simulation for the hypothesis $M_{\PH} = 1000\GeVcc$, which has the largest mass resolution
of the signal points studied and hence yields the most conservative limits. The limits on $N_X$
are close to 3.0 at most masses, as one would expect from Poisson statistics with zero observed signal,
but are larger at mass points near the masses of dilepton candidates observed in the
data. Since the fitted background levels under the signal peak are extremely small, the limits are not
expected to depend on the background shape, and indeed using the alternatives described in
Section~\ref{sec:bkgndest} give negligible changes in the results.  This figure also shows the 95\%
CL expected limit band. Except near the \Z~resonance, the a priori predictions of the
background normalisation are very small, so the expected limit is close to 3.0 and the expected limit
band is extremely narrow; the median value of the expected limit is in fact equal to 3.0 everywhere.

The expected number of signal dilepton candidates passing the selection cuts can be
expressed as:
\begin{eqnarray}\label{eqnlimit}
N_X  & = & 2 \Lumi \epsilon_{1} \sigma \BR \left[1 + \BR(\epsilon_{2}/{\epsilon_{1}} - 1)\right]
\end{eqnarray}
where $\Lumi$ is the integrated luminosity, $\epsilon_{(1,2)}$ are the efficiencies defined in
Section~\ref{sec:effi}, $\sigma$ is the production cross section of the heavy resonance decaying to \X\X,
and \BR is the branching fraction for the decay $X \rightarrow \lp\lm$. This expression takes into
account that either one or both \X~bosons in an event may decay to the chosen lepton species, and that, as
mentioned in Section~\ref{sec:effi}, the efficiency to select such an \X~boson is slightly different in the
two cases. Using this equation, the likelihood function can be expressed in terms of $\sigma\BR$, thus
allowing upper limits to be placed on this quantity.
Since $N_X$ in Eq.~(\ref{eqnlimit}) depends not only on $\sigma\BR$, but also on \BR, the
upper limits depend on the assumed value of \BR.  However, the factor
$(\epsilon_{2} / \epsilon_{1} - 1)$ is in practice always positive or very small. Hence if one
assumes infinitesimally small \BR when calculating the limits on $\sigma\BR$, such that the factor in
square brackets in Eq.~(\ref{eqnlimit}) is equal to 1, the resulting limits will be valid, and in some
cases conservative, for any value of \BR.

\begin{figure*}[hbtp]
\centering
\includegraphics[width=0.49\textwidth]{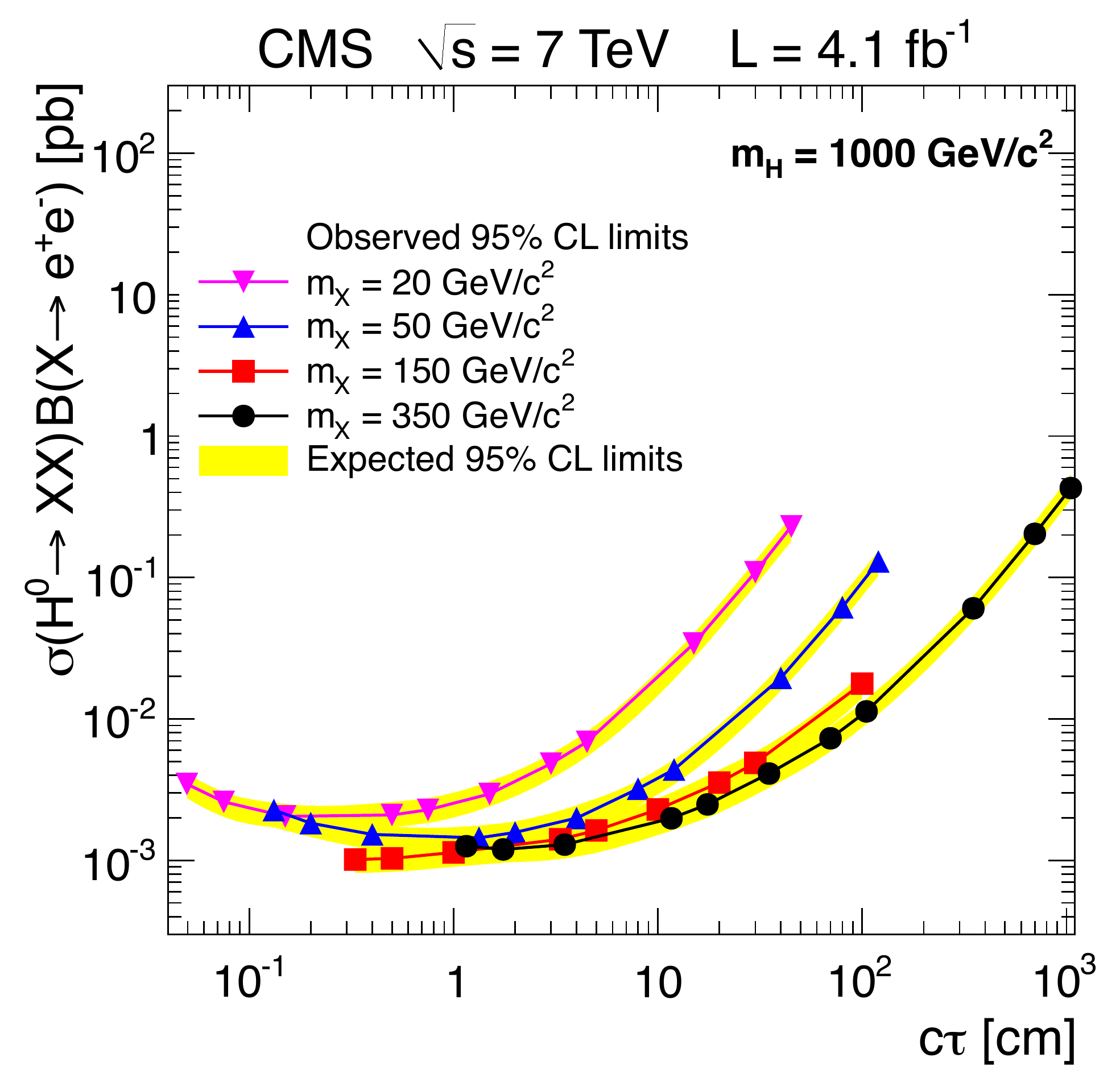}
\includegraphics[width=0.49\textwidth]{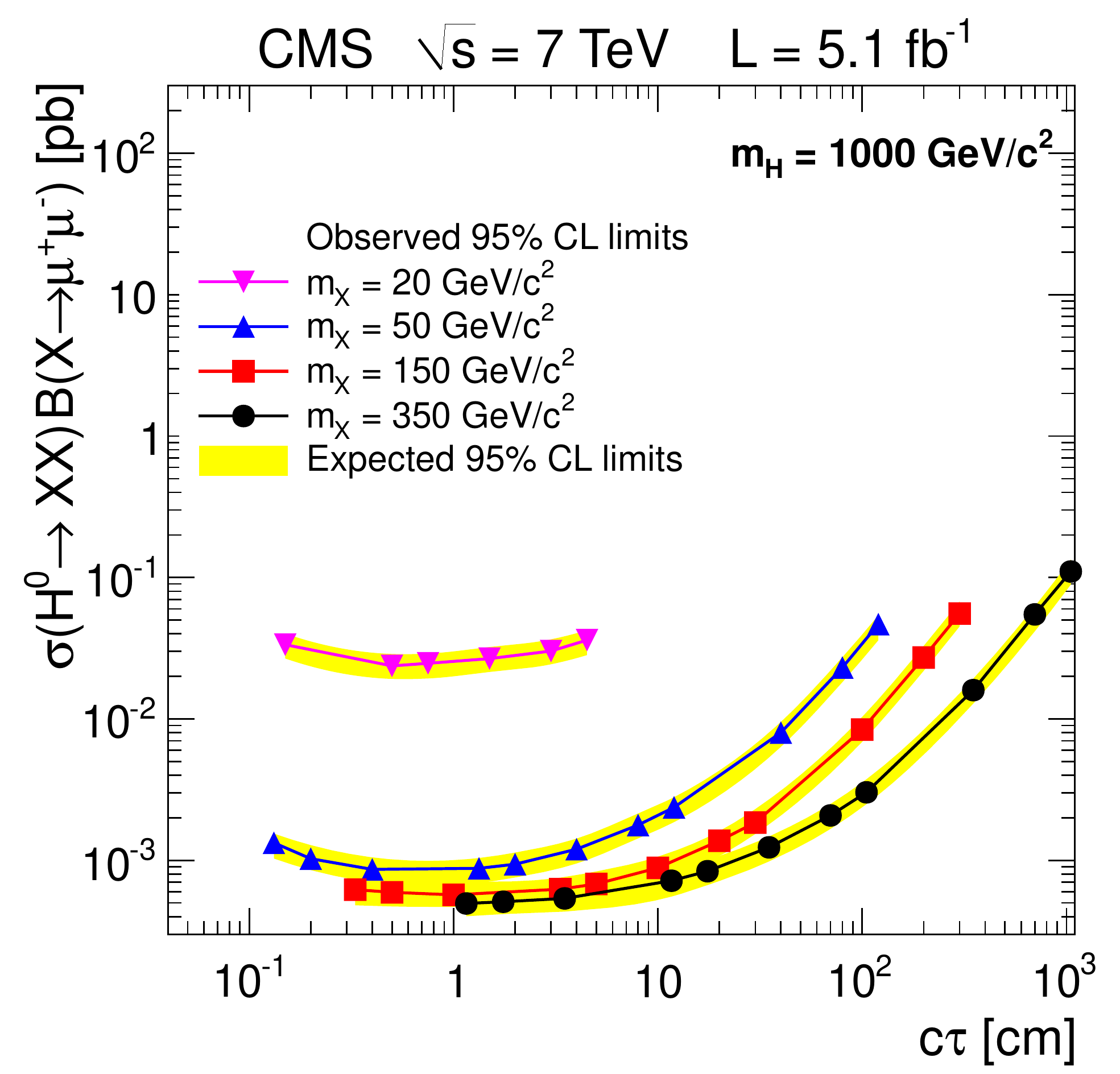}
\caption{The 95\% CL upper limits on $\sigma \BR$ for the
electron (left) and muon channel (right) for a \Higgs~mass of 1000\GeVcc. Narrow yellow shaded bands show the
95\% quantiles for the expected limits.}
\label{fig:FinLimits1000}
\end{figure*}

\begin{figure*}[hbtp]
\centering
\includegraphics[width=0.49\textwidth]{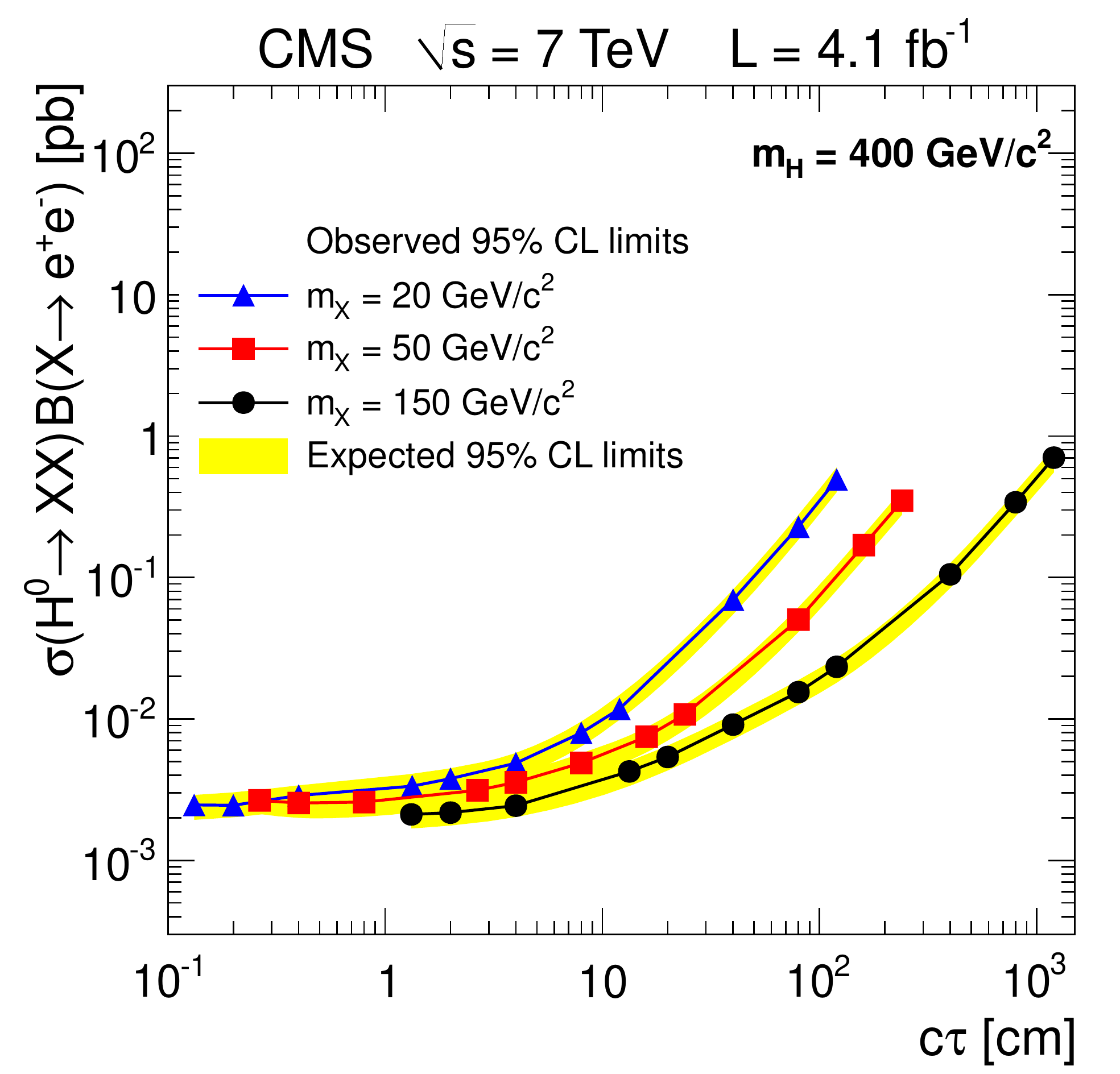}
\includegraphics[width=0.49\textwidth]{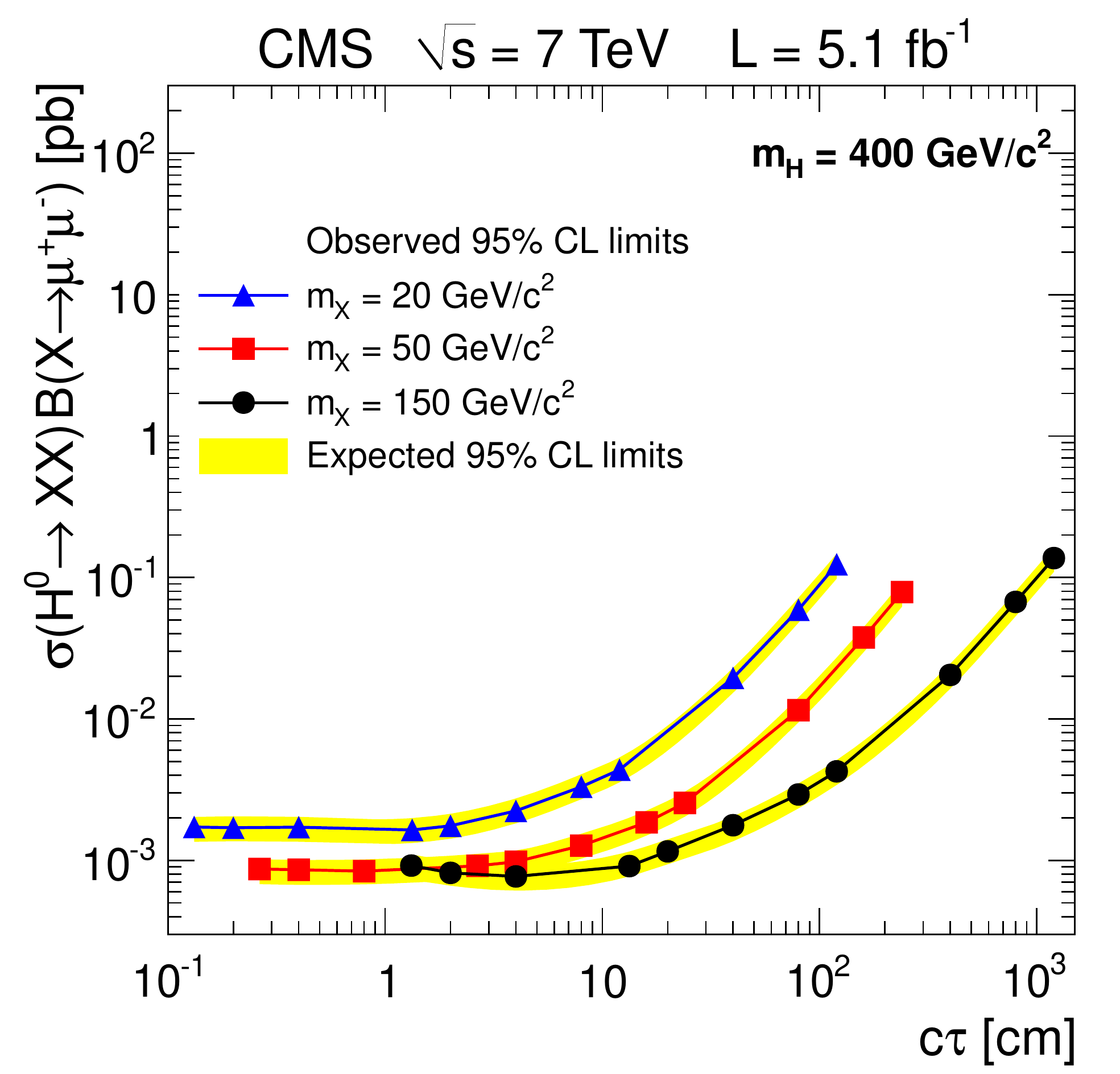}
\caption{The 95\% CL upper limits on $\sigma \BR$ for the
electron (left) and muon channel (right) for a \Higgs~mass of 400\GeVcc. Narrow yellow shaded bands show the
95\% quantiles for the expected limits.}
\label{fig:FinLimits400}
\end{figure*}

\begin{figure*}[hbtp]
\centering
\includegraphics[width=0.49\textwidth]{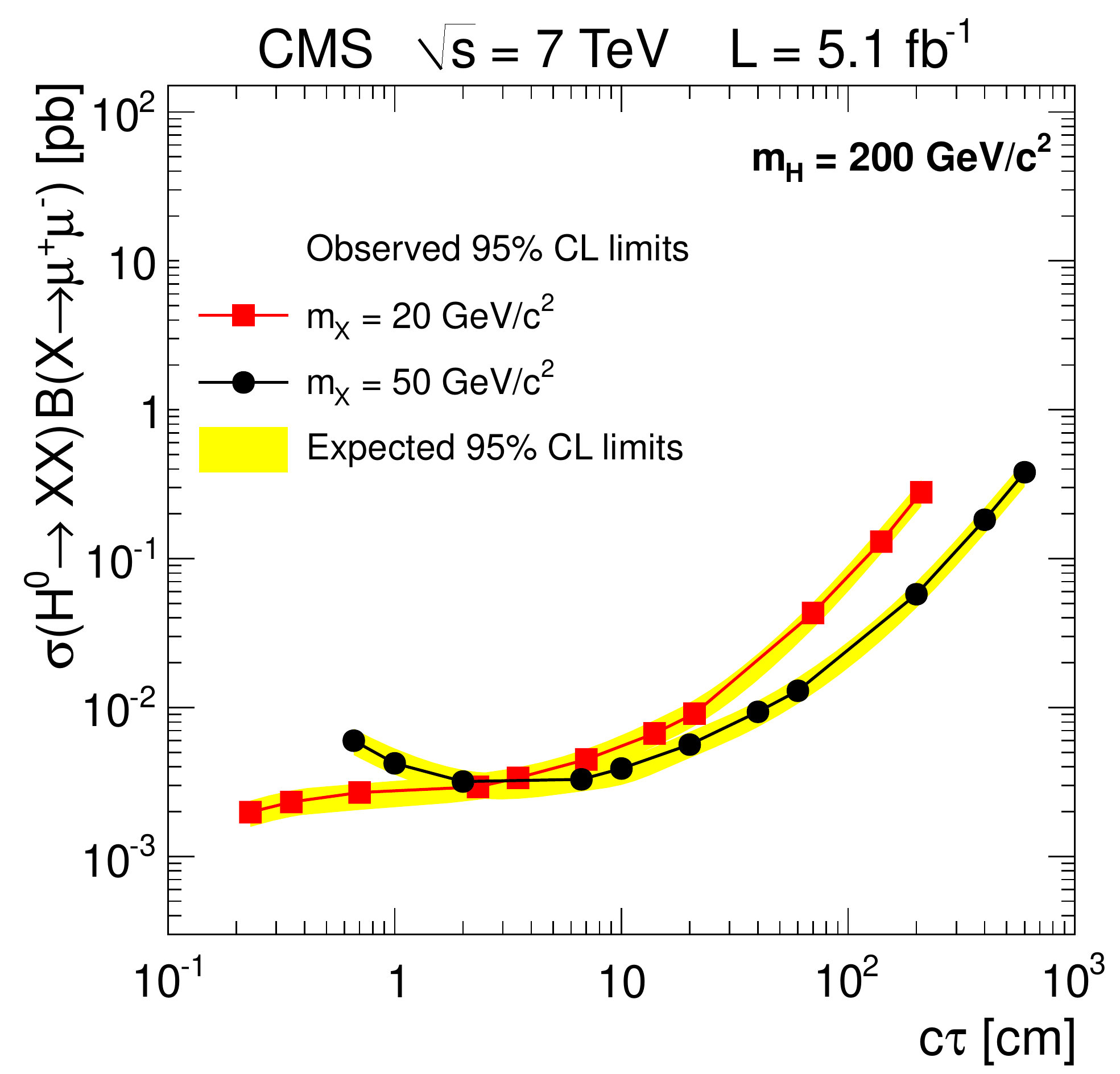}
\includegraphics[width=0.49\textwidth]{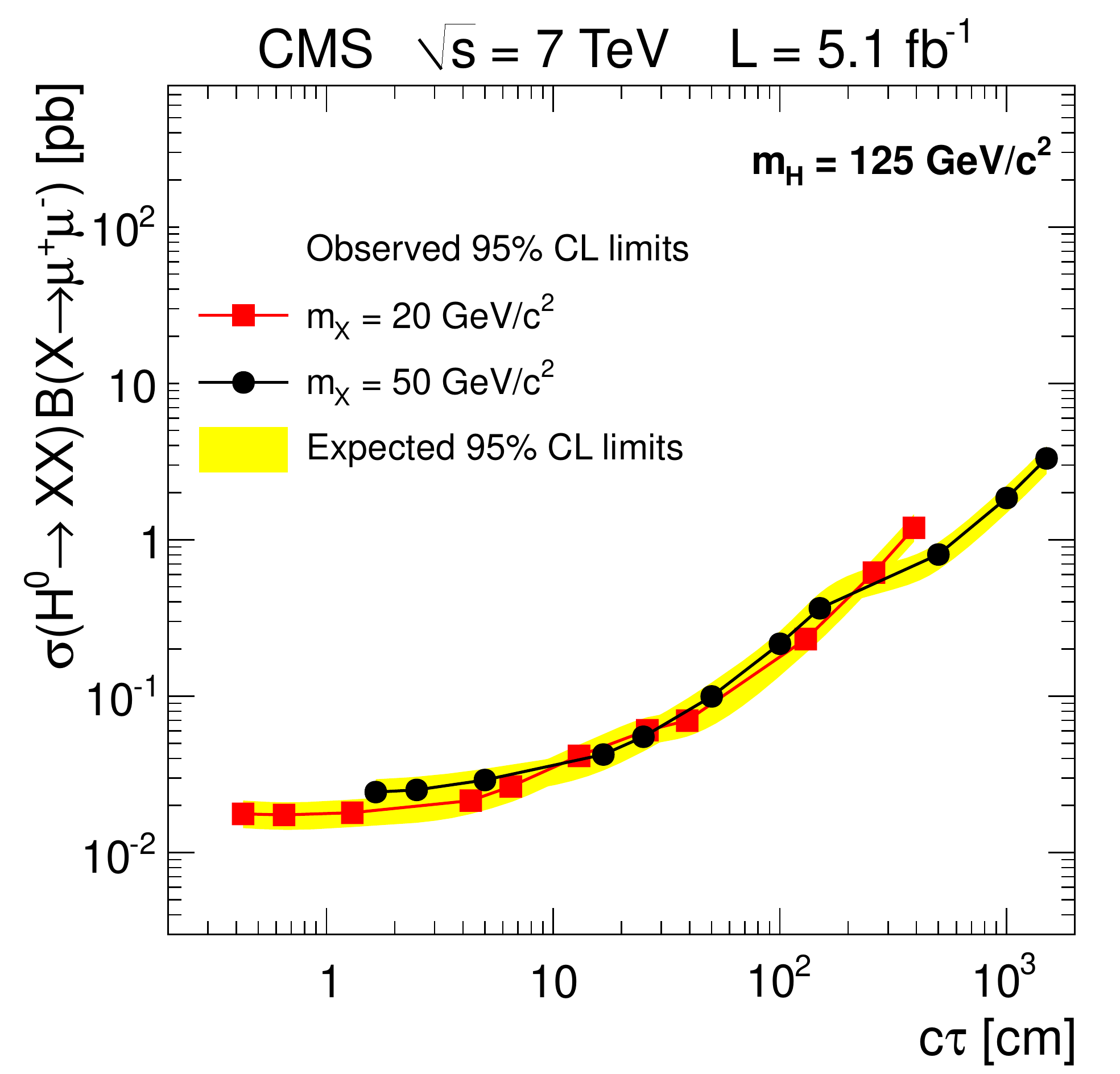}
\caption{The 95\% CL upper limits on $\sigma \BR$ for the
muon channel for a \Higgs~mass of 200\GeVcc (left) and 125\GeVcc (right). Narrow yellow shaded bands show the
95\% quantiles for the expected limits.}
\label{fig:FinLimits125}
\end{figure*}

For each
combination of the \Higgs~and \X~boson masses that are modelled, and for
a range of X boson lifetimes, the 95\% CL upper limits on $\sigma\BR$ are calculated.
The observed limits are shown in
Figs.~\ref{fig:FinLimits1000}--\ref{fig:FinLimits125}.
(No results are shown for $M_{\Higgs}\le 200\GeVcc$ in the electron channel, since
the high trigger thresholds result in a very low signal efficiency.)
Note that for the muon channel in the $M_{\Higgs} = 1000\GeVcc$,
$M_{\X} = 20\GeVcc$ case, the efficiency is significantly reduced
because the muons are produced very close together, which causes
trigger inefficiencies.
Since the observed dilepton candidates do not have masses close to those of the \X~bosons
considered in these plots, they have no effect on the limits. The bands show the 95\% quantile for the
expected limits.

For \Higgs~or \X~boson masses other than those plotted in
Figs.~\ref{fig:FinLimits1000} -- \ref{fig:FinLimits125}, exact limits are not computed, since no simulated
signal samples are available with which to determine the signal selection efficiency. However, since the
observed limits appear to be monotonic functions of the \Higgs~and \X~boson masses, one can infer
approximate limits for other masses, provided the latter lie within the range of those shown in the
figures. For example, for $M_{\Higgs}=1000\GeVcc$, it should be safe to assume that the limits for
$M_X=170\GeVcc$ would be at least as good as the weaker of the limits for $M_X=150\GeVcc$ and $M_X=350\GeVcc$. However, for \X~bosons that are close in mass to the candidates seen in data, the limits
would be worse than this. For these particular masses, Fig.~\ref{fig:FinLimitsNum} gives an indication of
the factors by which the limits would be degraded.

In an alternative signal model, in which $\Higgs\rightarrow\X\Y$, where
the \Y~boson does not decay to dileptons, the expected number of signal candidates $N_X$ passing the selection cuts
can be expressed as $N_X = \Lumi\epsilon_{1}'\sigma\BR$, where $\epsilon_{1}'$ is the efficiency to select
signal candidates, in this scenario. If the \X and \Y bosons have identical masses, then
$\epsilon_{1}' = \epsilon_{1}$, and comparison with Eq.~(\ref{eqnlimit}) shows that the limits would be
a factor of 2 worse than those presented in Figs.~\ref{fig:FinLimits1000} -- \ref{fig:FinLimits125}.

The limits quoted above are for \Higgs bosons produced through
gluon-gluon fusion.  If the \Higgs bosons were instead produced by the sum of all standard model
production mechanisms, their momentum spectra would be different. For $M_{\Higgs}=125\GeVcc$, the selection
efficiency would then be larger by a factor of approximately 1.18 (1.08) for $M_X=20$~(50)\GeVcc and there would be a corresponding
improvement in the limits.

If the initial resonance were a \Zprime with spin 1 instead of a \Higgs boson, the
acceptance for a dilepton pair to pass the \pt and rapidity selection cuts would
be slightly different, mainly because \Zprime are produced by $\cPq\cPaq$
annihilation, whilst \Higgs~are produced predominantly through gluon-gluon
fusion.
If the \Zprime is to decay to a pair of long-lived spin~0 particles, these
cannot be identical if CP is to be conserved, but they are assumed in what
follows to have equal mass. Their angular distributions would differ depending
on whether they were fundamental spin~0 bosons or spin~$\frac{1}{2}$ `hidden
valley' quarks that have hadronised in the dark sector into spin~0 bosons
\cite{Han:2007ae}. Studies with simulated events
show that the change in the acceptance for these two \Zprime models relative to the original \Higgs~model is less than 3\% for a \Zprime mass of 1000\GeVcc, less than 11\% for a 400\GeVcc \Zprime mass, and below 25\% (in
the muon channel) for a 200\GeVcc \Zprime mass.
Thus the
limits quoted above on the \Higgs~model will be approximately valid for these
two \Zprime models as well, for \Zprime masses of 200\GeVcc or more.

\section{Summary}

A search for long-lived neutral particles, \X, produced in pp collisions at $\sqrt{s} = 7$\TeV and decaying
to either $\Pep\Pem$ or $\mu^+\mu^-$ has been performed. In the $\Pep\Pem$ channel 4 candidates are observed, of which
2 are in the \Z mass region, and in the $\mu^+\mu^-$ channel no candidates are observed. These results are  consistent
with standard model expectations and are used to derive upper limits on the product of cross section times
branching fraction for a Higgs boson, in the mass range 200 -- 1000\GeVcc, decaying into a pair of \X~bosons, in the
mass range 20--350\GeV, which each decay to $\Pep\Pem$ and $\mu^+\mu^-$. The limits are typically in the range
0.7--10\unit{fb}, for \X~bosons with lifetimes in the range $0.1 < c\tau < 200\unit{cm}$.  For a Higgs mass of 125\GeVcc,
the corresponding limits are in the range 10--100\unit{fb}. These are the most stringent limits in these channels to date.

\section*{Acknowledgements}
We congratulate our colleagues in the CERN accelerator departments for the excellent performance of the LHC
and thank the technical and administrative staffs at CERN and at other CMS institutes for their contributions
to the success of the CMS effort. In addition, we gratefully acknowledge the computing centres and personnel
of the Worldwide LHC Computing Grid for delivering so effectively the computing infrastructure essential to
our analyses. Finally, we acknowledge the enduring support for the construction and operation of the LHC and
the CMS detector provided by the following funding agencies: BMWF and FWF (Austria); FNRS and FWO (Belgium);
CNPq, CAPES, FAPERJ, and FAPESP (Brazil); MEYS (Bulgaria); CERN; CAS, MoST, and NSFC (China); COLCIENCIAS
(Colombia); MSES (Croatia); RPF (Cyprus); MoER, SF0690030s09 and ERDF (Estonia); Academy of Finland, MEC, and
HIP (Finland); CEA and CNRS/IN2P3 (France); BMBF, DFG, and HGF (Germany); GSRT (Greece); OTKA and NKTH
(Hungary); DAE and DST (India); IPM (Iran); SFI (Ireland); INFN (Italy); NRF and WCU (Korea); LAS (Lithuania);
CINVESTAV, CONACYT, SEP, and UASLP-FAI (Mexico); MSI (New Zealand); PAEC (Pakistan); MSHE and NSC (Poland);
FCT (Portugal); JINR (Armenia, Belarus, Georgia, Ukraine, Uzbekistan); MON, RosAtom, RAS and RFBR (Russia);
MSTD (Serbia); SEIDI and CPAN (Spain); Swiss Funding Agencies (Switzerland); NSC (Taipei); ThEP, IPST and
NECTEC (Thailand); TUBITAK and TAEK (Turkey); NASU (Ukraine); STFC (United Kingdom); DOE and NSF (USA).

\bibliography{auto_generated}   

\providecommand{\href}[2]{#2}\begingroup\raggedright\begin{thebibliography}{10}%
\makeatletter
\providecommand{\hrefCMSnoop }[0]{\@secondoftwo}%
\makeatother
\providecommand{\doi}{\texttt{doi:}\begingroup \urlstyle{tt}\Url}

\bibitem{Hewett:2004nw}
J.~L. Hewett\hrefCMSnoop {} { {et~al.}, ``{Signatures of long-lived gluinos in
  split supersymmetry}'',} \textit{ JHEP} \textbf{ 09} (2004) 070,
  \href{http://dx.doi.org/10.1088/1126-6708/2004/09/070}{\doi{10.1088/1126-6708/2004/09/070}},
  \href{http://www.arXiv.org/abs/hep-ph/0408248}{\texttt{
  arXiv:hep-ph/0408248}}.

\bibitem{Barbier:2004ez}
R.~Barbier\hrefCMSnoop {} { {et~al.}, ``{R-parity violating supersymmetry}'',}
  \textit{ Phys. Rept.} \textbf{ 420} (2005) 1,
  \href{http://dx.doi.org/10.1016/j.physrep.2005.08.006}{\doi{10.1016/j.physrep.2005.08.006}},
\href{http://www.arXiv.org/abs/hep-ph/0406039}{\texttt{ arXiv:hep-ph/0406039}}.

\bibitem{Han:2007ae}
T.~Han\hrefCMSnoop {} { {et~al.}, ``{Phenomenology of hidden valleys at hadron
  colliders}'',} \textit{ JHEP} \textbf{ 07} (2008) 008,
  \href{http://dx.doi.org/10.1088/1126-6708/2008/07/008}{\doi{10.1088/1126-6708/2008/07/008}},
  \href{http://www.arXiv.org/abs/0712.2041}{\texttt{ arXiv:0712.2041}}.

\bibitem{Basso:2008iv}
L.~Basso\hrefCMSnoop {} { {et~al.}, ``{Phenomenology of the minimal B-L
  extension of the standard model: Z' and neutrinos}'',} \textit{ Phys. Rev. D}
  \textbf{ 80} (2009) 055030,
  \href{http://dx.doi.org/10.1103/PhysRevD.80.055030}{\doi{10.1103/PhysRevD.80.055030}},
  \href{http://www.arXiv.org/abs/0812.4313}{\texttt{ arXiv:0812.4313}}.

\bibitem{Strassler:2006ri}
\hrefCMSnoop {} {M.~J. Strassler and K.~M. Zurek, ``{Discovering the Higgs
  through highly-displaced vertices}'',} \textit{ Phys. Lett. B} \textbf{ 661}
  (2008) 263,
  \href{http://dx.doi.org/10.1016/j.physletb.2008.02.008}{\doi{10.1016/j.physletb.2008.02.008}},
  \href{http://www.arXiv.org/abs/hep-ph/0605193}{\texttt{
  arXiv:hep-ph/0605193}}.

\bibitem{Abazov:2006as}
\hrefCMSnoop {} {{ D0} Collaboration, ``{Search for neutral, long-lived
  particles decaying into two muons in $p \bar{p}$ collisions at $\sqrt{s}$ =
  1.96-TeV}'',} \textit{ Phys. Rev. Lett.} \textbf{ 97} (2006) 161802,
  \href{http://dx.doi.org/10.1103/PhysRevLett.97.161802}{\doi{10.1103/PhysRevLett.97.161802}},
  \href{http://www.arXiv.org/abs/hep-ex/0607028}{\texttt{
  arXiv:hep-ex/0607028}}.

\bibitem{Abazov:2008zm}
\hrefCMSnoop {} {{ D0} Collaboration, ``{Search for long-lived particles
  decaying into electron or photon pairs with the D0 detector}'',} \textit{
  Phys. Rev. Lett.} \textbf{ 101} (2008) 111802,
  \href{http://dx.doi.org/10.1103/PhysRevLett.101.111802}{\doi{10.1103/PhysRevLett.101.111802}},
  \href{http://www.arXiv.org/abs/0806.2223}{\texttt{ arXiv:0806.2223}}.

\bibitem{ATLAS:2012av}
\hrefCMSnoop {} {{ ATLAS} Collaboration, ``{Search for a light Higgs boson
  decaying to long-lived weakly-interacting particles in proton-proton
  collisions at $\sqrt{s} = 7$ TeV with the ATLAS detector}'',} \textit{ Phys.
  Rev. Lett.} \textbf{ 108} (2012) 251801,
  \href{http://dx.doi.org/10.1103/PhysRevLett.108.251801}{\doi{10.1103/PhysRevLett.108.251801}},
\href{http://www.arXiv.org/abs/1203.1303}{\texttt{ arXiv:1203.1303}}.

\bibitem{Aad:2011zb}
\hrefCMSnoop {} {{ ATLAS} Collaboration, ``{Search for displaced vertices
  arising from decays of new heavy particles in 7 TeV pp collisions at
  ATLAS}'',} \textit{ Phys. Lett. B} \textbf{ 707} (2012) 478,
  \href{http://dx.doi.org/10.1016/j.physletb.2011.12.057}{\doi{10.1016/j.physletb.2011.12.057}},
\href{http://www.arXiv.org/abs/1109.2242}{\texttt{ arXiv:1109.2242}}.

\bibitem{JINST}
\hrefCMSnoop {} {{ CMS} Collaboration, ``The {CMS} experiment at the {CERN}
  {LHC}'',} \textit{ JINST} \textbf{ 03} (2008) S08004,
\href{http://dx.doi.org/10.1088/1748-0221/3/08/S08004}{\doi{10.1088/1748-0221/3/08/S08004}}.

\bibitem{Khachatryan:2010pw}
\hrefCMSnoop {} {{ CMS} Collaboration, ``{CMS tracking performance results from
  early LHC operation}'',} \textit{ Eur. Phys. J. C} \textbf{ 70} (2010) 1165,
  \href{http://dx.doi.org/10.1140/epjc/s10052-010-1491-3}{\doi{10.1140/epjc/s10052-010-1491-3}},
  \href{http://www.arXiv.org/abs/1007.1988}{\texttt{ arXiv:1007.1988}}.

\bibitem{PYTHIA}
\hrefCMSnoop {} {T.~Sj{\"o}strand, S.~Mrenna, and P.~Z. Skands, ``{PYTHIA} 6.4
  physics and manual'',} \textit{ JHEP} \textbf{ 05} (2006) 576,
  \href{http://dx.doi.org/10.1088/1126-6708/2006/05/026}{\doi{10.1088/1126-6708/2006/05/026}},
\href{http://www.arXiv.org/abs/hep-ph/0603175}{\texttt{ arXiv:hep-ph/0603175}}.

\bibitem{GEANT4}
\hrefCMSnoop {} {{ GEANT4} Collaboration, ``{GEANT4}---a simulation toolkit'',}
  \textit{ Nucl. Instrum. Meth. A} \textbf{ 506} (2003) 250,
\href{http://dx.doi.org/10.1016/S0168-9002(03)01368-8}{\doi{10.1016/S0168-9002(03)01368-8}}.

\bibitem{cms:lumi11WinterUpdate}
\href {http://cdsweb.cern.ch/record/1434360} {{ CMS} Collaboration, ``{Absolute
  Calibration of the Luminosity Measurement at CMS: Winter 2012 Update}'',} CMS
  Physics Analysis Summary CMS-PAS-SMP-12-008, (2012).

\bibitem{Bourilkov:2006cj}
\hrefCMSnoop {} {D.~Bourilkov, R.~C. Group, and M.~R. Whalley, ``{LHAPDF: PDF
  use from the Tevatron to the LHC}'',} (2006).
  \href{http://www.arXiv.org/abs/hep-ph/0605240}{\texttt{
  arXiv:hep-ph/0605240}}.

\bibitem{pas-trk-10-003}
\href {http://cdsweb.cern.ch/record/1279138} {{ CMS} Collaboration, ``{Studies
  of tracker material}'',} CMS Physics Analysis Summary CMS-PAS-TRK-10-003,
  (2010).

\bibitem{Khachatryan:2010xn}
\hrefCMSnoop {} {{ CMS} Collaboration, ``{Measurements of inclusive W and Z
  cross sections in pp collisions at $\sqrt{s} = 7$ TeV}'',} \textit{ JHEP}
  \textbf{ 01} (2011) 080,
  \href{http://dx.doi.org/10.1007/JHEP01(2011)080}{\doi{10.1007/JHEP01(2011)080}},
\href{http://www.arXiv.org/abs/1012.2466}{\texttt{ arXiv:1012.2466}}.

\bibitem{Read:2002hq}
\hrefCMSnoop {} {A.~L. Read, ``{Presentation of search results: the $CL_s$
  technique}'',} \textit{ J. Phys. G} \textbf{ 28} (2002) 2693,
\href{http://dx.doi.org/10.1088/0954-3899/28/10/313}{\doi{10.1088/0954-3899/28/10/313}}.

\bibitem{Junk:1999kv}
\hrefCMSnoop {} {T.~Junk, ``{Confidence level computation for combining
  searches with small statistics}'',} \textit{ Nucl. Instrum. Meth. A} \textbf{
  434} (1999) 435,
  \href{http://dx.doi.org/10.1016/S0168-9002(99)00498-2}{\doi{10.1016/S0168-9002(99)00498-2}},
\href{http://www.arXiv.org/abs/hep-ex/9902006}{\texttt{ arXiv:hep-ex/9902006}}.

\end{thebibliography}\endgroup

\cleardoublepage \appendix\section{The CMS Collaboration \label{app:collab}}\begin{sloppypar}\hyphenpenalty=5000\widowpenalty=500\clubpenalty=5000\textbf{Yerevan Physics Institute,  Yerevan,  Armenia}\\*[0pt]
S.~Chatrchyan, V.~Khachatryan, A.M.~Sirunyan, A.~Tumasyan
\vskip\cmsinstskip
\textbf{Institut f\"{u}r Hochenergiephysik der OeAW,  Wien,  Austria}\\*[0pt]
W.~Adam, E.~Aguilo, T.~Bergauer, M.~Dragicevic, J.~Er\"{o}, C.~Fabjan\cmsAuthorMark{1}, M.~Friedl, R.~Fr\"{u}hwirth\cmsAuthorMark{1}, V.M.~Ghete, N.~H\"{o}rmann, J.~Hrubec, M.~Jeitler\cmsAuthorMark{1}, W.~Kiesenhofer, V.~Kn\"{u}nz, M.~Krammer\cmsAuthorMark{1}, I.~Kr\"{a}tschmer, D.~Liko, I.~Mikulec, M.~Pernicka$^{\textrm{\dag}}$, D.~Rabady\cmsAuthorMark{2}, B.~Rahbaran, C.~Rohringer, H.~Rohringer, R.~Sch\"{o}fbeck, J.~Strauss, A.~Taurok, W.~Waltenberger, C.-E.~Wulz\cmsAuthorMark{1}
% ...
\vskip\cmsinstskip
\dag:~Deceased\\
1:~~Also at Vienna University of Technology, Vienna, Austria\\
2:~~Also at CERN, European Organization for Nuclear Research, Geneva, Switzerland\\
% ...

\end{sloppypar}
\end{document}